# Selecting Steady and Transient Photocurrent Response in BaTiO$_3$ Films

*F. Liu[1], I. Fina[2,3,*], D. Gutiérrez[1], G. Radaelli[4], R. Bertacco[4], J. Fontcuberta[1,*]*

[1]Institut de Ciència de Materials de Barcelona (ICMAB-CSIC), Campus UAB, Bellaterra 08193, Catalonia, Spain
[2]Max Planck Institute of Microstructure Physics, Weinberg 2, Halle, Germany
[3]University of Warwick, Department of Physics, Coventry CV4 7AL, United Kingdom
[4]LNESS Center - Dipartimento di Fisica del Politecnico di Milano, Como 22100, Italy

Correspondance to: fontcuberta@icmab.cat; ignasifinamartinez@gmail.com

The ferroelectric polarization and short-circuit photocurrent in BaTiO$_3$ thin films have been studied for different contact configurations that allow to measure the photoresponse and polarization under the presence of large or negligible imprint field. It is found that in all cases, the direction of the photocurrent is dictated by the depolarizing field and ultimately by the film polarization, with a negligible contribution of the imprint electric field. However, dramatic differences are found in their time-dependent photoresponse. Whereas in presence of imprint, steady photocurrents are observed under suitable illumination, transient photocurrents are generated in absence of imprint. It is argued that this distinct behavior is determined by the different Schottky barrier height at electrodes which thus offers a simple way to tune the film photoresponse. These findings could be exploited for electro-optic read-out and writing of ferroelectric memories.

# 1. Introduction

In semiconductors, the absorption of a photon of suitable energy can generate an electron-hole (e-h) pair. To electrically detect the e-h pair it is necessary to split it and drain the electron and hole charges thus generating current (so-called photocurrent) that is proportional to the number of photogenerated carriers. Drain can occur due to the presence of an internal electric field. Ferroelectrics have recently attracted a renewed interest due to the observation of above-bandgap open circuit photovoltage,[1,2] or significant change on their transport properties driven by the ferroelectric polarization state or light.[3–9] The photoresponse of ferroelectric materials is primarily determined by its polarization state, the Schottky barriers at interfaces and the concomitant built-in electric fields, and the so-called bulk photovoltaic effect (BPE), which is related to the non-centrosymmetric character of the ferroelectric materials.[10,11] Assuming homogeneous light illumination, the charge-draining electric field can be either generated by a built-in potential steaming from: differences between the work functions of electrodes, polarization gradients, or asymmetric distribution of defects, all of them being at the origin of the so-called imprint electric field ($E_{imp}$), or due to unscreened polarization and the concomitant depolarizing electric field ($E_{dep}$).[5,12–17] Alternatively, it is known[18] and recently reported in thin films,[19] that ferroelectrics may display BPE because, even in absence of any built-in electric field $E_{imp}$ or $E_{dep}$, photocarriers are asymmetrically scattered and generate a net charge flow.[11,13,14]

Exhaustive studies have been done in order to infer the correlation between $E_{imp}$ or $E_{dep}$ electric field and the generated photocurrent.[5,12–16] However, the relation between them and the generated photocurrent remains unclear because separation of both contributions is challenging. Indeed conflicting results are found on the literature where $E_{imp}$,[12,13] $E_{dep}$,[6,20,21] or both[22,23] are claimed to be the driving force for observed photocurrent. On the other hand, the time response of photocurrents in photoferroelectrics is primarily dictated by the generation and recombination rates of photocarriers giving rise to transient responses, typically in the ms range, that we call here "steady" photoresponse because the equilibrium state is reached fast.[24-26] However, in some cases, transient photoresponses with intriguing orders of magnitude slower responses, ranging from few seconds to thousands of seconds, of undisclosed origin, have been reported.[6,13,21,27-32] In the present article, we disentangle the contributions of $E_{imp}$ and $E_{dep}$ to the short-circuit photocurrent measured in BaTiO$_3$ (BTO) thin films, and we show that transient or steady photocurrent can be selectively obtained. A simple contact configuration, which can be generalized to any ferroelectric thin film, allows us to distinguish between $E_{imp}$ and $E_{dep}$ contributions to the photocurrent. It turns out that the ultimate parameter determining the photocurrent (magnitude and sign) is the depolarizing field. This allows univocally inferring the polar state of the layer from the measured photocurrent. Beyond this finding, we also show that, by selecting the appropriate contact configuration: asymmetric or symmetric, on the very same film, steady or transient photoresponses are selectively obtained. We argue that this dramatic difference is determined by Schottky barriers. Possible implications for data writing and reading in ferroelectric memories are addressed.

## 2. Results

BaTiO$_3$(150 nm)/La$_{2/3}$Sr$_{1/3}$MnO$_3$(50 nm) (LSMO) bilayers were grown in a single process by pulsed laser deposition on (001) SrTiO$_3$ (STO) substrates. After cooling down to room temperature, 20 nm thick platinum top electrodes of $60 \times 60$ μm$^2$ separated each by about 15 μm, were deposited ex-situ on the BTO surface by RF-sputtering, by using a mask allowing to deposit about 100 contacts simultaneously. See experimental sections for details.

In **Figure 1**a,b, we show the two different contact configurations used for electric measurements, that we named bottom-top (b-t) and top-top (t-t), respectively. In the b-t configuration (Figure 1a) the Pt top electrode is contacted and the bottom LSMO electrode is grounded. This contact configuration corresponds to a single capacitor with asymmetric electrodes (Pt and LSMO). We have defined the positive sign of the current (labeled "*j*" in Figure 1a,b) as that where positive carriers flow from Pt to LSMO (ground), as indicated by the arrow in Figure 1a. In the t-t configuration (Figure 1b), two top Pt electrodes are contacted, resulting in a nominally symmetric contact configuration. In this t-t configuration one of the Pt electrodes is grounded and the other is contacted. As before, positive current implies positive charges flowing towards the ground (arrow in Figure 1b).

The *P-E* loops recorded for both configurations are shown in Figure 1c,d. In b-t configuration (Figure 1c) a positive or negative voltage applied between top-Pt and bottom-LSMO results in saturated polarization pointing towards LSMO or Pt respectively, as sketched in the insets of the Figure 1c. In t-t configuration (Figure 1d) a positive voltage applied between two top adjacent Pt electrodes produces a pointing-down polarization ($P > 0$) in the capacitor where positive voltage is applied and pointing-up polarization ($P < 0$) in the grounded capacitor, and vice versa when the polarity of the poling field is reversed, as sketched in the insets of the Figure 1d. From now on, positive poling indicates a positive voltage applied to the non-grounded electrode.

Comparison of *P-E* data in Figure 1c and 1d reveals the profound impact of the contact configuration on the measured polarization loops. In Figure 1c, it is obvious that the *P-E* loop recorded in b-t configuration is shifted towards the right, revealing the presence of large imprint ($E_{imp} \approx -150$ kV/cm). Notice that $E_{imp} < 0$ implies that $E_{imp}$ points from LSMO towards Pt. The opening of the branches of the *P(E)* at negative voltages (at about -300 kV/cm) is a signature of leakage whereas the aperture of the loop at the largest positive voltages is consequence of the non-saturation of the film.[33–35] In the t-t configuration (Figure 1d), the almost symmetric *P-E* loop indicates that $E_{imp}$ is virtually absent. This does not imply that $E_{imp}$ has disappeared, but its strength is overruled by the series connection of unscreened ferroelectric capacitors. Indeed, if in t-t configuration $E_{imp}$ would act, head-to-head and tail-to-tail ferroelectric capacitors, short-circuited via the common bottom electrode, would occur which would have a large electrostatic energy payload. Note also here the presence of the aperture of *P-E* loops near the highest applied voltage, consequence of the leakage current.

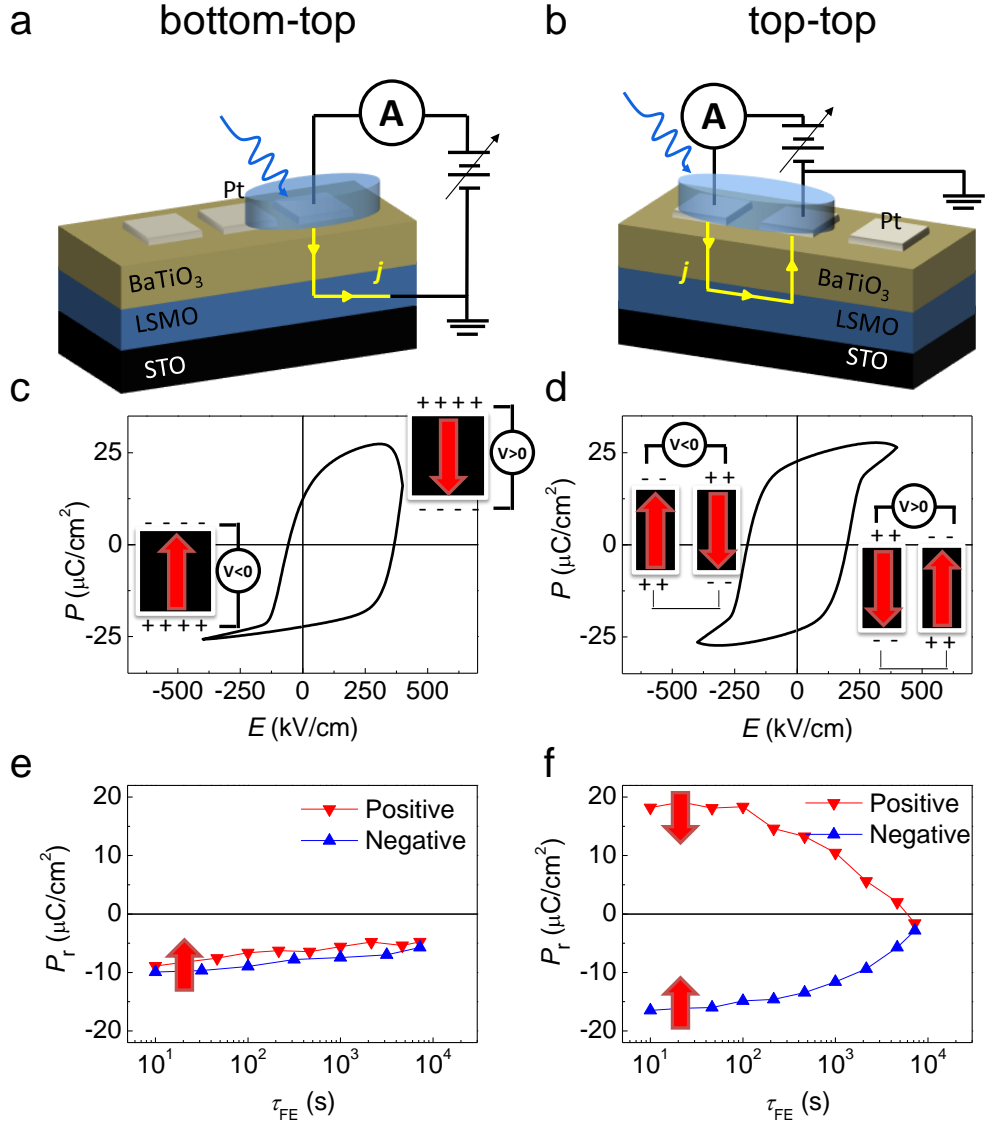

**Figure 1.** (a,b) Electric contacts configuration used for **b-t** (a) and **t-t** (b) measurements. Arrows indicate the positive sign of the current for positive applied voltage. (c,d) Dynamic ferroelectric hysteresis *P-E* loops recorded in dark at 200 Hz for **b-t** (c) and **t-t** (d) configurations. Sketched are the relations between applied voltage and ferroelectric polarization direction for both configurations and prepoling voltage sign. (e,f) Ferroelectric retention measurements obtained after pre-poling (with "positive" or "negative" voltage pulse, as labeled in the figure) the sample in a given direction and measuring the switchable charge, after a delay time $\tau_{FE}$. Red arrows indicate the ferroelectric polarization direction.

The presence of $E_{imp}$ and $E_{dep}$ is well visible in the retention measurements displayed in Figure 1e,f, where we show the remnant polarization $P_r$ as a function of delay time between the prepoling pulse and the measurement pulse ($\tau_{FE}$). For b-t configuration (Figure 1e), $P_r$ is found to be very similar irrespective on the sign of the prepoling pulse (positive or negative). Indeed, right at the very first measurement ($\tau_{FE} = 10$ s), the polarization is found to be already around -10 µC/cm$^2$, irrespectively of the sign of the prepoling voltage. This is the consequence of the presence of $E_{imp}$, that produces a back-switch of the ferroelectric polarization after a short time (< 10 s) towards the most stable configuration ($P$ pointing up). Moreover, one can observe in Figure 1e that there is a gradual decrease of the absolute $P_r$ with $\tau_{FE}$: more than 50% of the initial polarization ($\tau_{FE} \approx 10$ s) is retained after $\tau_{FE} \approx 10^4$ s. This observation signals the presence of the coexisting depolarizing field $E_{dep}$.

In the configuration t-t, it can be clearly appreciated (Figure 1f) that the sign of the remnant polarization is dictated by the sign of the poling field; this is fully consistent with the observed symmetric $P$-$E$ loops (Figure 1d) and the corresponding virtual absence of $E_{imp}$. In contrast, the presence of $E_{dep}$ is more visible, being responsible for the gradual loss of polarization with time. In this t-t configuration, the effect of a minor residual $E_{imp}$ on $P_r$ is evidenced by the slightly negative $P_r$ observed at $\tau_{FE} \approx 10^4$ s, irrespectively on the initial poling direction. One notices in Figure 1f that $P_r$ is reduced by 50% at $\tau_{FE} \approx 10^3$ s compared to the initial value. Therefore, in t-t configuration the polarization decays faster than in b-t, where, as mentioned, $P_r$ is reduced by less than 50% at $10^4$ s.

In **Figure 2**, we show the dependence on time ($t$) and illumination delay time ($\tau_{light}$) of the short-circuit photocurrent $j(t, \tau_{light})$ for both contact configurations. Short-circuit photocurrent is measured for various $\tau_{light}$ (from 10 s up to 2 h) after a prepoling voltage pulse (a triangular pulse up to ±400 kV/cm applied for 2.5 ms).

Figure 2a and 2c display the $j(t, \tau_{light})$ data collected for configuration b-t after negative and positive poling pulses, respectively. It is observed that when light is switched on, indicated by the small red arrows in the figures, the current gradually increases until reaching a steady state that persists during illumination and only washes out after switching off the illumination. This photoresponse is similar to that observed in common pn-junctions. However, a striking observation is the fact that, irrespectively on the poling voltage sign, the photocurrent is always positive.

In Figure 2b and 2d, we show the corresponding data recorded in t-t configuration after positive and negative voltage poling, respectively. In a back-to-back configuration of two photodiodes, a negligible photoresponse should be expected, because both photodiode contributions would cancel out. Here, a radically different photocurrent response is observed, indicating that the switchable polarization rather than the contact asymmetry, is the driving force for the observed effects. Indeed, the sign of $j(t, \tau_{light})$ is always dictated by the poling direction and reverses when reversing the poling voltage. Here, $j_{max}$ is also found to be dependent on $\tau_{light}$, but in sharp contrast with the photoresponse of b-t, in t-t configuration the photocurrent rapidly decays with time during illumination and it displays a change of sign at longer times. On the other hand, BPE does not play a major role on the measured photoresponse[11] because we have observed that the open-circuit voltage is much smaller than the BTO bandgap and that the photocurrent $j(t, \tau_{light})$ measured in similar but thicker films is reduced (not shown).

The first insight into the distinct photoresponse behavior of the same film under different contact configurations can be obtained from the comparison of the photocurrent data (Figure 2a-d) and the retention experiments (Figure 1e,f). It is obvious that the sign of $j(t, \tau_{light})$ univocally depends on the sign of the polarization. To emphasize this observation we include in Figure 2a-d big red arrows indicating the direction of the observed polarization.

We turn now towards the observed time-dependencies of $j(t, \tau_{light})$. In Figure 2e,f we plot $j_{max}$ vs. $\tau_{light}$ for both contact configurations. First, we note that in b-t measurements (Figure 2e), the extrapolated time $\tau_{light}$ to reach about 50% of $j_{max}$ is much longer than $10^4$ s whereas in the t-t (Figure 2f), it is shorter than $10^3$ s. A similarly different time-dependence was already observed in the retention measurements of Figure 1e,f. Indeed, if one compares Figure 1e,f and Figure 2e,f, their resemblance is rather impressive (except for the opposite sign). Therefore, one can conclude that there is an intimate relationship between the polarization retention and the measured short-circuit photocurrent.

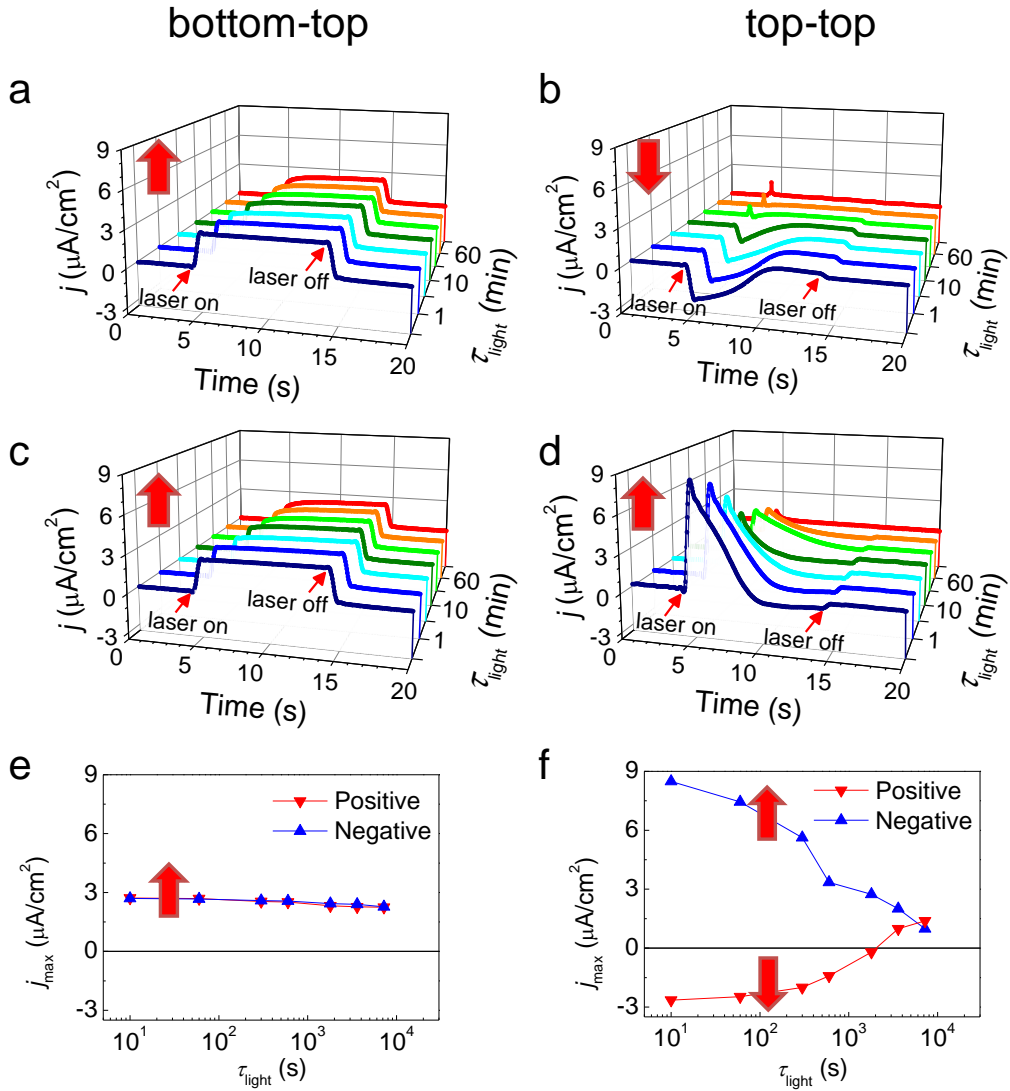

**Figure 2.** Time dependence of short-circuit photocurrent for different illumination delay times ($\tau_{light}$ = 10 s, 1 min, 5 min, 10 min, 30 min, 60 min and 120 min). (a,c) Data collected for **b-t** configuration after a positive and negative pre-poling, respectively. (b,d) Data collected for **t-t** configuration after a positive and negative pre-poling. The big red arrows indicate the majority polarization direction. (e,f) Dependencies of the maximum photocurrent ($j_{max}$) on $\tau_{light}$ for **b-t** and **t-t** configurations, respectively.

On the other hand, it is well-visible that when using t-t configuration, the measured photocurrent has a time dependence that is strikingly different than that of photoactive structures based on semiconducting materials. Just after switching on the illumination, the current rapidly rises up to $j_{max}$ and it gradually decreases (roughly by 50% in 1 – 3 s) until reaching a steady state. We also note that the current density at the maximum $j_{max}$ lowers with increasing $\tau_{light}$, roughly by 50% at 10 min ($\approx$ 10$^3$ s). We stress that the time constant of the measuring circuit is much shorter ($\approx$ 10 µs), thus implying that the measured $j(t)$ in both configurations is a genuine photoresponse. However, it does not result from the $E_{dep}$ identified in polarization measurements, as shown by data in Figure 1e,f, because the depolarization time is several orders of magnitude longer. The dependence of $j_{max}$ on $\tau_{light}$ and the physics of the transient peak [$j(t)$] will be discussed below. Before proceeding with this discussion, we stress here that the observed photocurrent is a genuine photovoltaic effect rather than a spurious photoresponse, that could be originated by an inhomogeneous illumination or pyrocurrents. We performed similar experiments using laser of similar power but with photons of lower energy (green) and no photoresponse was observed (Figure S1, Supporting Information).

## 3. Discussion

Now we will discuss: i) the fact that the sign of $P_r$ determines $j_{max}$, ii) the correlation between the magnitude of both parameters, and iii) the distinct time dependence of the photocurrent $j(t)$ depending on the measurement configuration. Regarding point i), we first note that that the sign of the generated photocurrent is at odds with the sign of the ferroelectric $E_{imp}$. Indeed, as mentioned above, in b-t configuration the $E_{imp}$ points from bottom LSMO towards top Pt, and thus, if $E_{imp}$ was relevant, the photocurrent should flow in the same direction in the measuring circuit (Figure 1a), that is: negative, according to the used sign convention (see the sketches in Figure S2.1 and S2.2, Supporting Information). This is contrary to the observations (Figure 2a,c). Therefore we disregard a determinant role of the $E_{imp}$ field in the measured photoresponse. Instead, the direction of $j_{max}$ is highly correlated with $P_r$ as inferred from the close similarity between the time dependence of the polarization retention $P_r(\tau_{FE})$ (Figure 1e,f) and the dependence of $j_{max}$ on $\tau_{light}$ (Figure 2e,f), that has prompted a more detailed consideration. In **Figure 3**a,b we plot the dependence of $j_{max}(\tau_{light})$ on $P_r(\tau_{FE})$ extracted from Figure 1e,f and Figure 2e,f, for both configurations. From Figure 3a, where data for b-t are plotted, the narrow range of $P_r$ variation experimentally accessible does not allow to extract any relevant conclusion on the impact of $P_r$ on $j_{max}$. However, for t-t, as shown in Figure 3b, $j_{max}$ vs. $P_r$ data display a remarkably simple linear dependence over a wide range of $P_r$ values including its change of sign. This linear relation can result from the presence of $E_{dep}$, i.e. $j = \sigma E_{dep}$, where film conductivity $\sigma$ contains the photoinduced carrier density, and the depolarizing field $E_{dep}$ which itself is proportional to $P_r$. It is known that $E_{dep} = -\alpha \dfrac{P}{\varepsilon}$,[36,37] where $\varepsilon$ is the dielectric permittivity of the ferroelectric and $\alpha$ accounts for the fraction of the unscreened polarization; therefore, $j = -\dfrac{\sigma \alpha}{\varepsilon} P$. On the other hand, it is known that in defective ferroelectric thin films, as it is the present case, a fraction of ferroelectric domains could be pinned and not switchable (or harder to switch) under the application of an electric field.[38,39] Therefore at remanence $P = P_r + P_{pin}$, where $P_r$ accounts for the switchable polarization and $P_{pin}$ for the pinned polarization. It results that $j$ must show a linear dependence on $P_r$ as follows:

$$j = -\dfrac{\sigma \alpha}{\varepsilon} P_r - \dfrac{\sigma \alpha}{\varepsilon} P_{pin} \qquad (1)$$

For simplicity we have assumed in Equation (1) that the permittivity and conductivity of pinned and switchable ferroelectric domains are identical. Therefore, being $j_{max} \propto j$, Equation (1) predicts a linear $j_{max}(P_r)$ dependence as observed (Figure 3b). Accordingly, data of Figure 3b have been fitted to the Equation (1). Using the conductance of our film $\sigma \approx 1.5 \times 10^{-4}$ µS/cm, extracted from the slope of $j$-$E$ characteristic under the used illumination (Figure S3, Supporting Information) and $\varepsilon \approx 300\ \varepsilon_0$, extracted from the saturated *P-E* loops recorded (t-t configuration) under illumination (Figure S4, Supporting Information), we obtain: $\alpha \approx 0.03$ (well within the common order of magnitude[37]), $E_{dep} \approx 1.3$ kV/cm (for 1 µC/cm$^2$) and the polarization of the pinned domains is 4.8 µC/cm$^2$ (pointing towards the ground).

Some $j_{max}$ data points in Figure 3b, corresponding to the largest negative polarization, do not follow the predicted linear dependence. Within the scope of the model derived above, this implies that for this negative polarization the screening is less efficient. A possible reason could be that the available charges for screening are fewer and, accordingly, $E_{dep}$ increases. We have observed a similar linear $j_{max}(P_r)$ dependence and we have obtained similar $\alpha$ values in other STO/LSMO/BTO samples grown and contacted alike, see Figure S5 of the Supporting Information.

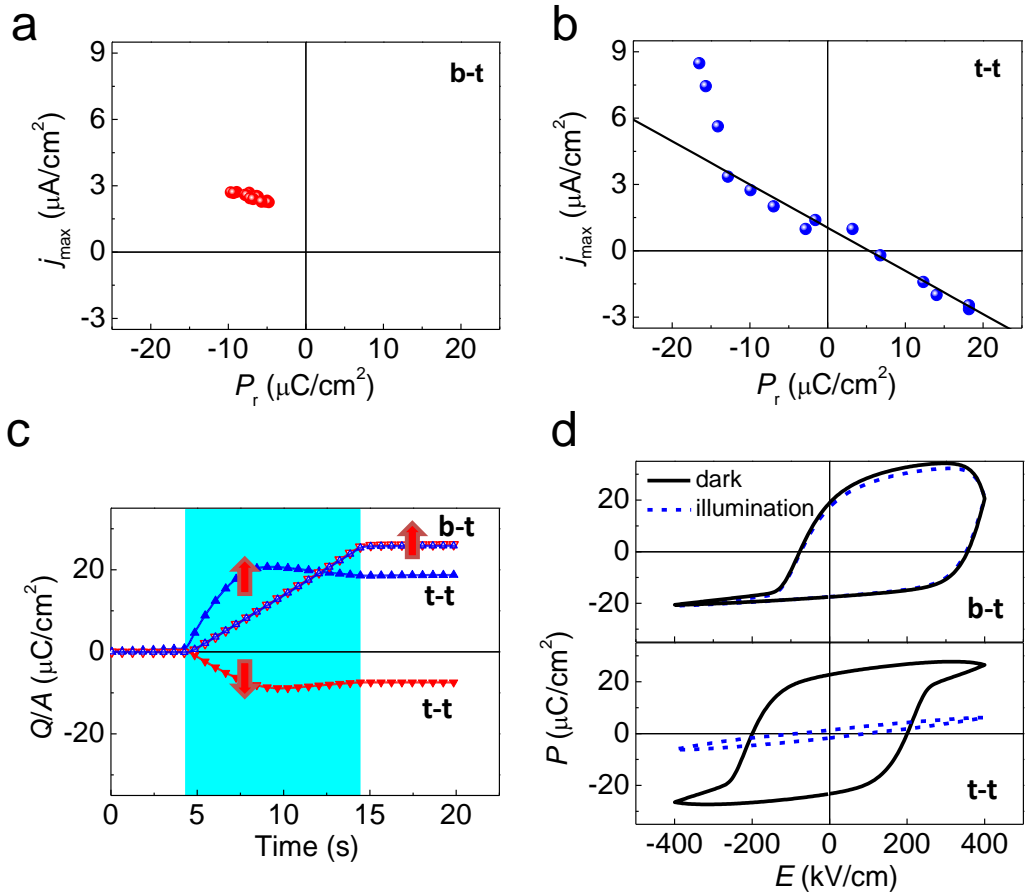

**Figure 3.** Dependence of the maximum of the photocurrent on remnant polarization determined from ferroelectric and photocurrent measurements shown in Figure 1 and 2 for (a) **b-t** and (b) **t-t** contact

configuration. Line through data points corresponds to the fitted line: $j_{max} = 1.0(1) \mu A/cm^2 + 0.2(1) s^{-1} P_r$ (c) Time dependence of the integrated charge flow [$Q = \int I(t) \, dt$] normalized to the contact area from the short-circuit photocurrent measurements shown in Figure 2 with a delay time of 10s for different polarization states and different contact configurations, as indicated. (d) *P-E* loops recorded in **b-t** (upper panel) and **t-t** (bottom panel) at 200 Hz in dark and under illumination.

Finally, we focus on the time dependence of the photocurrent observed in t-t and b-t configurations. In Figure 3c, we plot the time dependence of the charge that has flowed through the circuit normalized to the contact area (*Q/A*) obtained by integrating the photocurrent, at $\tau_{light} = 10$ s, measured for t-t and b-t configurations (from Figure 2a-d). It is observed that for b-t, the charge that has circulated increases linearly during illumination and only stops increasing when light is switched off. This effect obviously results from the fact that the photocurrent is virtually constant during illumination (Figure 2a,c) and thus there is not upper bound for the total amount of flowing charge. In the t-t measurement, in contrast, the integrated charge does not grow anymore after a few seconds, in spite the sample is still being illuminated. The integrated charge is bounded by $\approx +20$ μC/cm$^2$ and -10 μC/cm$^2$. To understand this distinct response, *P-E* loops have been recorded in both configurations in dark and under illumination (Figure 3d). It can be observed that in b-t [Figure 3d (upper panel)] the *P-E* loop recorded under light overlaps the one recorded in dark. In contrast, in t-t [Figure 3d (bottom panel)] the loop recorded under illumination shows a reduction of the remnant polarization by $\Delta P = 20$ μC/cm$^2$. This value coincides with the upper limit of *Q/A* (time) shown in Figure 3c. This observation strongly indicates that in t-t configuration the photogenerated charges screen the polarization, similar to early observations of Dimos and Warren[40–44] and Land and Peercy[32,45–48] in BTO and in doped PZT, respectively. At this point it may be worth to recall that Wurfel and Batra,[49] in their pioneering work on depolarizing fields, reported the opposite effect, that is an increase of polarization under illumination in metal/ferroelectric/semiconductor structures, which was attributed to photoinduced increase of carriers in the semiconductor electrode. The observed time scale for the transient photocurrent (few seconds), that is the time scale to build a space charge compensating the depolarizing field, is orders of magnitude shorter than the carrier drift time $\tau_d$. Indeed, for BTO, assuming a mobility $\mu_n \approx 1$ cm$^2$/(Vs), $E_{dep} \approx 2.3$ kV/cm and an electrode-to-electrode distance of about 15 μm, it turns out that $\tau_d < 0.1$ μs. This characteristic time is obviously much shorter than the experimental one (few seconds). These observations, which are in agreement with the reported time-dependence of photoinduced changes of surface potential in BTO crystals,[28,29] point that low-mobility charge species could be involved in the polarization screening process.

In short, both in b-t and t-t configurations the photocurrent is dictated by the polarization and the accompanying depolarizing field. The radical difference being that, under suitable illumination, in t-t the photocurrent charges are limited by the ferroelectric polarization whereas in b-t photocarriers are continuously generated and leave the sample. A schematic view of the charge distribution and electron energy representing the situation of distinct electrodes (b-t) (Pt/BTO/LSMO) and identical (t-t) (Pt/BTO/Pt) electrodes is shown in **Figure 4**a and 4b, respectively. The sketch is made under the assumption that BTO is an n-type semiconductor,[50] which can be related to the presence of oxygen vacancies, and a certain $E_{dep}$ exists across the BTO film. The nature of the screening charges carriers at BTO/Pt interface cannot be safely inferred from the available data. The space charge regions at each interface are schematically indicated (shadowed regions).

The situation of asymmetric interfaces (b-t) is depicted in Figure 4a. We assume that the Schottky barrier at the LSMO interfaces to be smaller than at the Pt side. This situation is analogous to that found in Nb:STO/BTO/Pt[51] where Nb:STO plays a similar role than LSMO here. This is consistent with the measured j-E curves that show a rectifying diode-like behavior characteristic of a metal/n-type semiconductor junction and displaying a larger conductance for a positive biased Pt electrode (Figure S6, Supporting Information). In the symmetric electrode structure (t-t case sketched in Figure 4b, where the contribution of the LSMO bottom electrode has been neglected) we notice that, even when the system is poled, the Schottky barrier height and the space charge width at each interface differ and these asymmetries change when reversing the polarization direction. Here both Schottky barriers at the interfaces act as blocking layers. Upon illumination photogenerated electron-hole pairs will be driven towards the ferroelectric surface by $E_{dep}$. As a consequence, the screening-charge distribution in the electrodes will be modified and the corresponding excess charge will flow from one electrode to the other across the measuring circuit. After some transient time, the situation will end up with: the total screening of $P_r$ and the concomitant suppression of $E_{dep}$, the flattening of the electron energy band (as indicated by the dashed band in the sketch of Figure 4b), and the suppression of the photocurrent. This total screening implies that the charge flow in the circuit will be limited by $P_r$, i.e. $Q/A \leq P_r$. This limit is indeed observed when photocurrent is measured in the t-t configuration (Figure 3c).

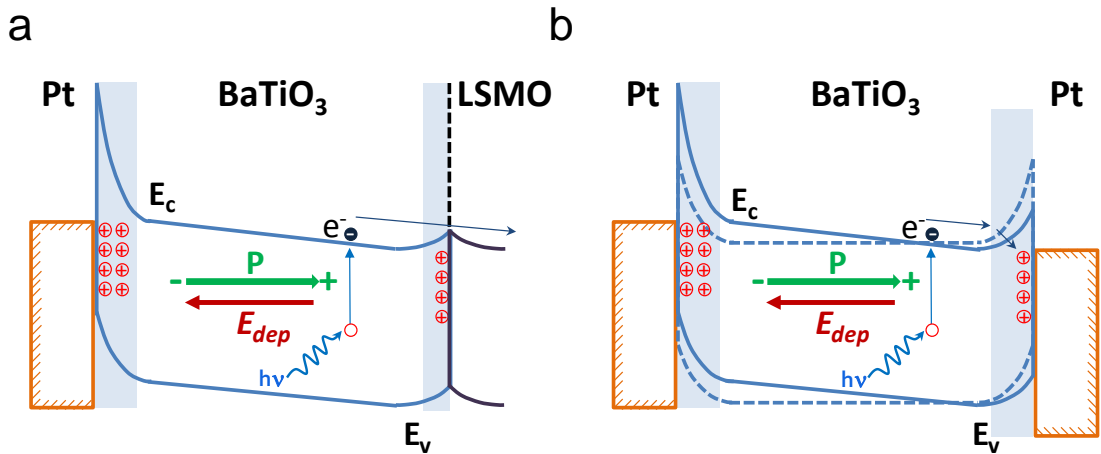

**Figure 4.** Sketch of the electronic energy band diagram in: (a) asymmetric **b-t** and (b) symmetric **t-t** configurations. Shadowed regions correspond to the space charge regions at each interface. The arrows indicate the trajectory of a photogenerated electron, flowing through the circuit in **b-t** (a) and being blocked at the interface in **t-t** (b). The blue dashed lines indicate the flattening of the electronic energy

band because of the total screening of $P_r$. The LSMO electrode is not included in the sketch because, being more conductive than the Pt/BTO interface, it does not play a relevant role. For simplicity, only photogenerated electrons are depicted and holes are omitted.

Before concluding we wish to comment on the assumption that the effective height of the Schottky barrier at the BTO/LSMO interface being relatively small and rather conducting. Carrier injection across Schottky barriers at high temperature, is commonly determined by thermionic injection, a process in which charges are thermally activated over an effective interface barrier of energy $W_B(V_A)$, where $V_A$ is the voltage across the device and $W_B(V_A) = W_0 - \frac{\sqrt{q^3 E_A}}{4\pi\varepsilon}$, with $E_A$ being the electric field at the interface and $W_0$ the zero field Schottky barrier. It follows that if $E_A$ increases the barrier high decreases. It is well known[52–54] that at interfaces between dielectrics and electrodes, the electric field can be much enhanced, particularly for defective dielectrics, being this enhancement more pronounced for relatively thick films.[52] Therefore, it is easily conceivable that the BTO/LSMO interface, with BTO film *ex-professo* made defective, could have a depressed barrier height. The observation of an imprint field at the BTO/LSMO interface pointing away from LSMO could also contribute to reduce any built-in potential at this interface.

Photoexcitation across the bandgap should create electron-hole pairs. In the description made above, for simplicity focus has been put on the photogenerated electron carriers, thus neglecting the role of holes. Holes would contribute also to screen the polarization. Their inclusion would not modify the picture we build. However, minority carriers are expected to have a smaller mobility and contribute less to the observed photocurrent.

# 4. Conclusion

Summarizing, we have used symmetric and asymmetric contact configurations in metal/ferroelectric/metal structures to disentangle the distinct role of imprint and depolarizing fields in the observed photoresponse. It is concluded that $E_{dep}$ is the actual driving field for the generated photocurrent which, accordingly, reverses its direction if the polarization is reversed irrespectively on an eventual presence of $E_{imp}$ which only imposes the direction of the polarization. Whereas the effect of the depolarizing field on the photocurrent magnitude and direction is found to be similar irrespectively on the contact configuration, dramatic differences are found in their time dependence photoresponses. Indeed, in presence of imprint, steady photocurrents are observed under suitable illumination, whereas transient photocurrents are generated in absence of an imprint. We have argued that this distinct behavior is controlled by the Schottky barriers at the electrodes, which thus offer a simple way to tune the film photoresponse. The observed univocal relation between photocurrent and polarization and its tuning by close-to-bandgap light exposure, in simple ferroelectric layers, may favor exploitation in optical writing and reading of ferroelectric memories. Indeed, pioneering attempts[41,55] to write polarization information used voltage biasing during illumination. The observed relevant role of the Schottky barrier demands further study of these effects with experiments involving electrodes with different conductors or controlled doping levels that can lead to enhanced responses. On the other hand, sensing polarization direction by photocurrents, demands that the electric field driving photogenerated carriers, shall be ruled by depolarizing field rather than other non-polarization dependent sources.[20] Here, we have shown that the contact configuration used allow to simultaneously alleviate these requirements in a simple way.

## 5. Experimental Section

BaTiO$_3$(150 nm)/La$_{2/3}$Sr$_{1/3}$MnO$_3$(50 nm) (LSMO) bilayers were grown in a single process by pulsed laser deposition on (001) SrTiO$_3$ (STO) substrates using a quadrupled Q-Switched Nd:YAG laser ($\lambda$ = 266 nm) with a fluence of 1.3 J cm$^{-2}$ for LSMO deposition process and 0.4 J cm$^{-2}$ for BTO deposition process and a repetition rate of 2 Hz. LSMO films were grown at a deposition temperature of 730 °C and an oxygen pressure of 0.22 Torr. The subsequent growth of BTO, performed at 640 °C and with an oxygen pressure of 0.02 Torr, was followed by 30 min annealing at 600 °C in a high oxygen pressure (760 Torr). After cooling down to room temperature, 20 nm thick platinum top electrodes of 60 × 60 µm$^2$ separated each by about 15 µm, were deposited ex-situ on the BTO surface by RF-sputtering, by using a mask allowing to deposit about 100 contacts simultaneously. X-ray diffraction experiments showed that the LSMO layer is epitaxial and fully strained whereas the BTO layer shows an elongated c-axis parameter with a near relaxed in-plane cell parameter (Figure S7, Supporting Information) as expected for a 150 nm BaTiO$_3$ film. Growth conditions have been selected to induce a significant expansion of the out-of-plane cell parameter, commonly associated to the presence of oxygen-stoichiometry deficit.[56]

Ferroelectric measurements were performed using a planar capacitor configuration. The polarization was evaluated by measuring the dynamic *P-E* hysteresis loops using a TFAnalyzer2000 (aixACCT Systems GmbH). Retention measurements were performed by poling the sample in a given direction with a triangular pulse of ±400 kV cm$^{-1}$ applied for 2.5 ms, to saturate it, and measuring the remnant polarization, using an identical pulse in the opposite direction, after a delay time ($\tau_{FE}$) ranging from 10 s to 10$^4$ s. The latter pulse switches the remnant polarization that is determined integrating the current through the measuring circuit. Further details on measurement protocols can be found in ref.[57]. As commonly observed when using ex-situ grown metal electrodes, $E_{imp}$ and $E_{dep}$ are found to be somehow depending on the particular contact (or pair of contacts) considered. We have tested many of them and selected the ones in which the effects of $E_{imp}$ on ferroelectric measurements are more apparent (in b-t configuration) and negligible (in t-t configuration).

Short-circuit photocurrent was measured by illuminating the STO/LSMO/BTO sample with blue laser of wavelength 405 nm feed by a CPX400SA DC power source (AimTTi Co.). Photoinduced current $j(t, \tau_{light})$ has been monitored as a function of time ($t$) and the time elapsed between ferroelectric (positive and negative) poling and switching on the illumination ($\tau_{light}$). We strength here that the used photons (blue; 3.06 eV) are of sub-bandgap energy (3.3 eV for BTO[58]). It is known that oxygen deficiencies (or other point defects) in $BaTiO_3$ introduce donor states, either shallow or deep[59–61] in the bandgap and thus significant photon absorption, even for sub-bandgap incoming photons, can be anticipated.[60,62,63] Indeed, light absorption experiments reveal an enhanced absorption at 3.06 eV (Figure S8, Supporting Information). The spot diameter is of 200 μm, safely illuminating homogeneously two adjacent electrodes, as sketched in Figure 1b.

**Acknowledgements**

Financial support by the Spanish Ministry Ministerio de Economia y Competitividad (MAT2011-29269-C03 and MAT2014-56063-C2-1-R), are acknowledged. Ignasi Fina acknowledges the Beatriu de Pinós postdoctoral scholarship (2011 BP-A 00220) from the Catalan Agency for Management of University and Research Grants (AGAUR-Generalitat de Catalunya). Fanmao Liu is financially supported by China Scholarship Council (CSC) with No. 201306020016. And we acknowledge C. Rinaldi and M. Cantoni for the growth of $BaTiO_3$ films.

# Supporting Information

**Selecting Steady and Transient Photocurrent Response in BaTiO$_3$ Films**

*F. Liu, I. Fina\*, D. Gutiérrez, G. Radaelli, R. Bertacco, J. Fontcuberta\**

**Supporting Information 1**

In Figure S1a and S1b the photoresponse measured, by using a green laser (532 nm wavelength) of similar power, in **b-t** and **t-t** configurations are shown. The absence of any measurable photocurrent indicates that heating or inhomogeneous illumination, are not at the origin of the effects observed using a blue laser.

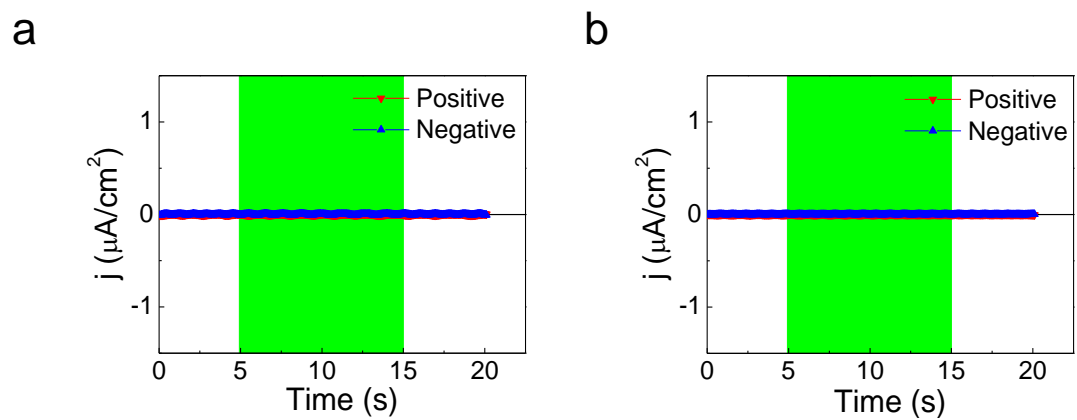

**Figure S1.** Short-circuit photocurrent dependence on time for **b-t** (a) and **t-t** (b) contact configurations under illumination with a green laser.

**Supporting Information 2**

In Figure S2.1, we show the shift in the *P-V* (equivalent *P-E*) loop (Figure S2.1a) and the correspondence between the imprint electric field in the structure (Figure S2.1b) and. Figure S2.2 shows the direction and the sign (in the ampere meter) of the generated photocurrent assuming an internal field pointing towards Pt as in Figure S2.1b.

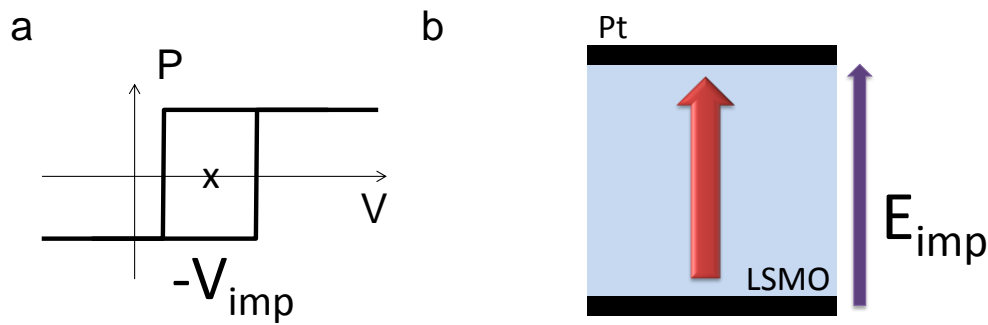

**Figure S2.1.** (a) Archetypical *P-V* loop for a ferroelectric material with strong imprint field. (b) Sketch of the direction of the polarization (red arrow) and the imprint field (purple arrow, $E_{imp}$) for an imprinted ferroelectric.

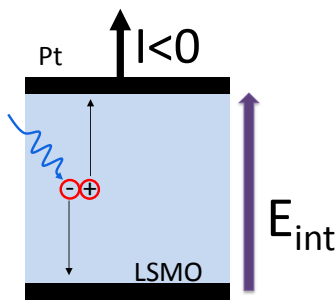

**Figure S2.2.** Sketch of the displacement of the carriers (electron-hole) photogenerated under the presence of an internal electric field ($E_{int}$). The directions of the carriers inside the material and the direction of the current in the circuit are indicated by the thin and thick black arrows, respectively.

**Supporting Information 3**

In Figure S3, current density vs. electric field characteristic under illumination recorded in **t-t**. From the slope of the fitted straight line (dashed) the conductivity $1.5 \times 10^{-4}$ µS/cm has been extracted.

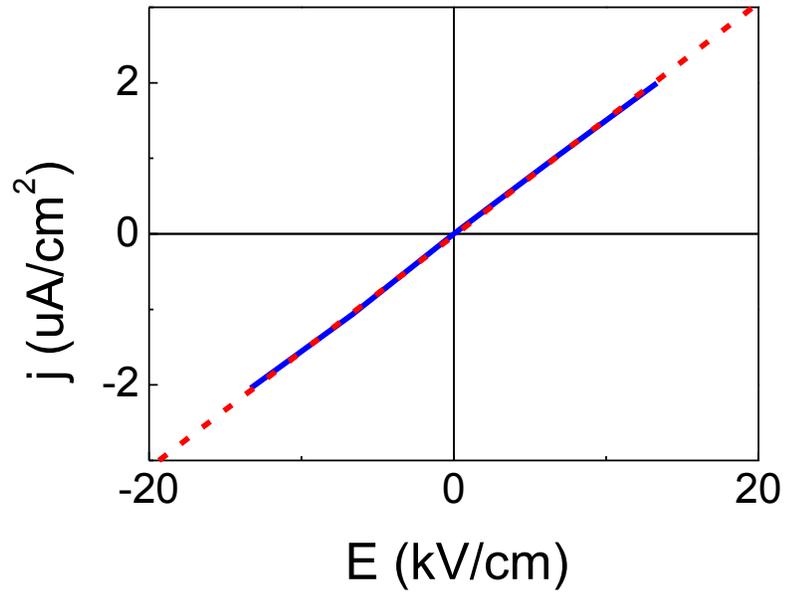

**Figure S3.** *j-E* characteristic in **t-t** configuration under light.

**Supporting Information 4**

In Figure S4, we show the permittivity $\varepsilon = dP/dE$ obtained from the derivative of the *P-E* loop in **t-t** configuration.

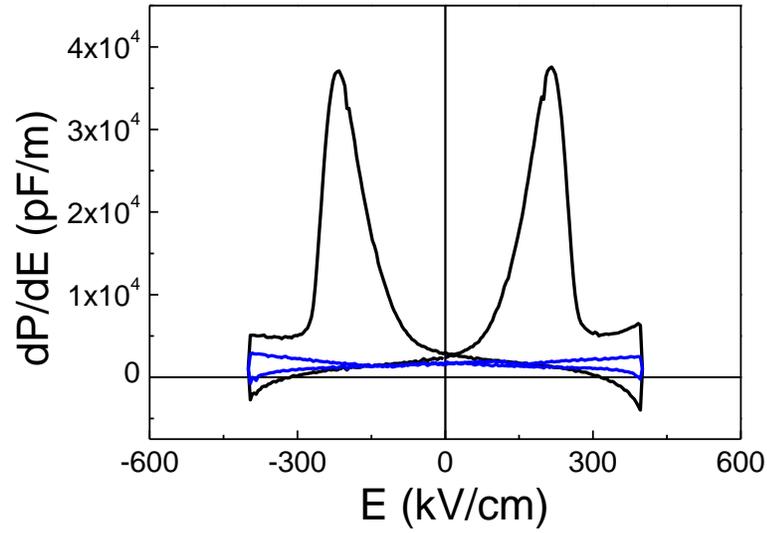

**Figure S4.** Derivative of the *P-E* loop in **t-t** configuration, recorded in dark (black line) and under light (blue line).

**Supporting Information 5**

In Figure S5, we plot $j_{max}$ vs. $P_r$ for **t-t** configuration for a different Pt/BTO/LSMO sample, where all layers have nominally the same thickness as the one described in the manuscript. From the fit of the data to Equation (1) of the manuscript, we obtain a fraction of unscreened polarization of $\alpha \approx 0.001$.

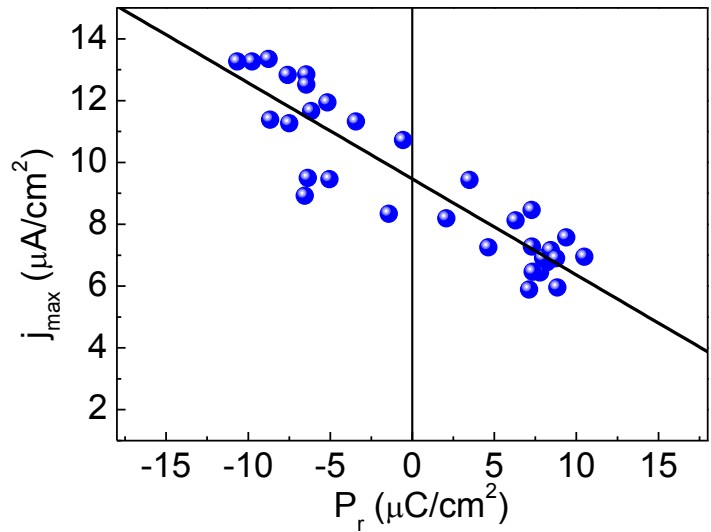

**Figure S5.** Dependence of the maximum of the photocurrent with remnant polarization extracted from separate ferroelectric and photocurrent measurements for **t-t** contact configuration for a different Pt/BTO/LSMO sample.

**Supporting Information 6**

In Figure S6, we show the *j-E* characteristic measured in the **b-t** configuration. In the inset, we indicate how sample is contacted, and the positive and negative voltage applied.

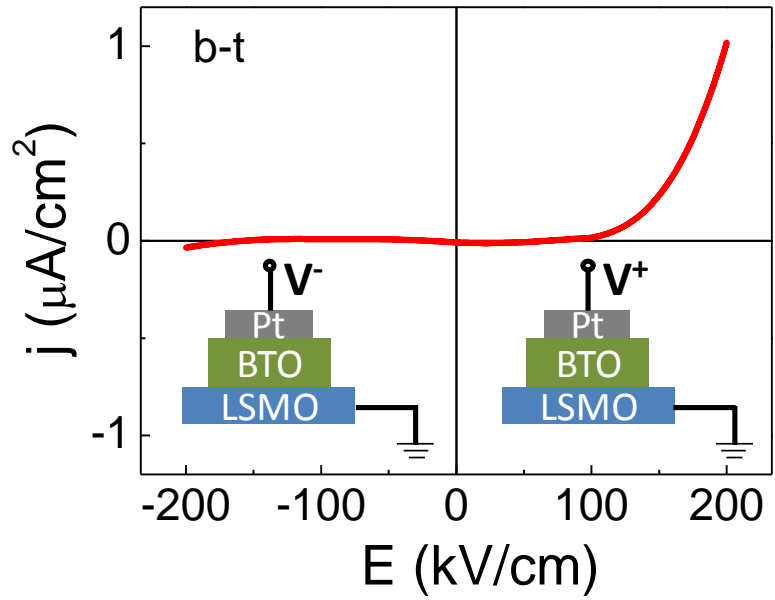

**Figure S6.** *j-E* characteristic recorded for the Pt/BTO/LSMO sample in **b-t** configuration in dark.

**Supporting Information 7**

In Figure S7, it is shown the XRD reciprocal space map (q-plot) around the STO (103) that reveals that the BTO film is epitaxial and in-plane relaxed [$a$ = 3.99(0) Å, the vertical line indicates the position of the corresponding bulk $a$ axis]. From the q-plot it can be inferred that the c-axis lattice parameter of BTO is 4.18(3) Å. From these values one obtains a unit cell volume of 66.59 Å$^3$, which is larger than its bulk value 64.38 Å$^3$. This difference is commonly found and attributed to oxygen vacancies.

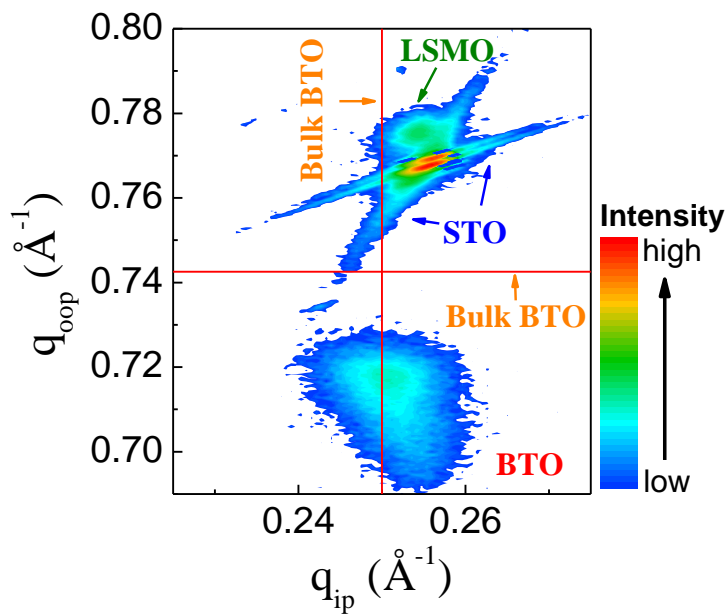

**Figure S7.** XRD reciprocal lattice maps around the (103) reflections of STO, LSMO, and BTO. The red lines mark the position of BTO (103) reflection in bulk.

**Supporting Information 8**

In Figure S8 we show the absorption spectra of a BTO/LSMO/STO(001) sample. For comparison data collected on a bare STO substrate is included. To emphasize changes in absorption close to the band edge, data are normalized at the corresponding values at energies well below the band edge. In Figure S8, it can be observed an enhanced absorption (see STO spectrum for comparison) at energies similar to the used blue laser light.

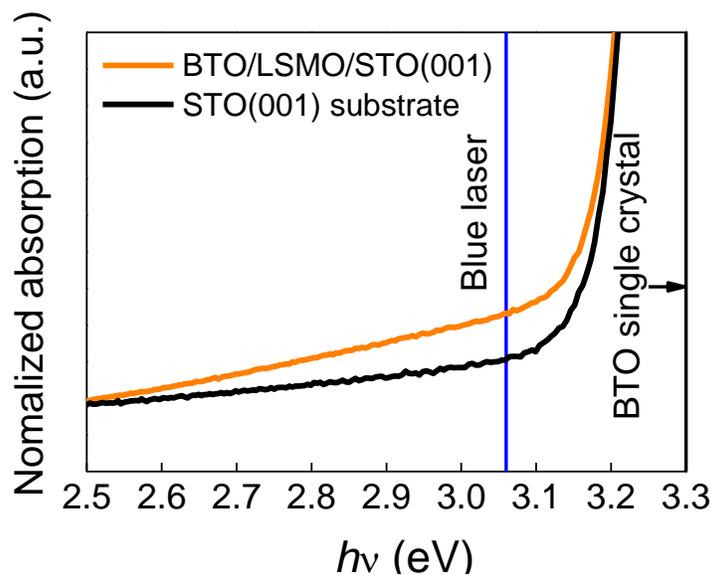

**Figure S8.** Absorption spectra normalized to their value at 2.5 eV of the BTO/LSMO/STO sample and of a bare STO (001) substrate.